# Dynamics of particles with cubic magnetic anisotropy in a viscous liquid

E. M. Gubanova[1], R. A. Rytov[2], N. A. Usov[1,2,3]

[1]*National Research Nuclear University "MEPhI", 115409, Moscow, Russia*
[2]*National University of Science and Technology «MISiS», 119049, Moscow, Russia*
[3]*Pushkov Institute of Terrestrial Magnetism, Ionosphere and Radio Wave Propagation, Russian Academy of Sciences, IZMIRAN, 108480, Troitsk, Moscow, Russia*

**Abstract** The specific absorption rate (SAR) of a dilute assembly of spherical iron nanoparticles with cubic anisotropy distributed in a viscous liquid is calculated using the solution of stochastic Landau - Lifshitz equation for unit magnetization vector and stochastic equations for multiple particle directors that specify the spatial orientation of the nanoparticle in a liquid. The viscous and magnetic magnetization reversal modes of particles are revealed at low and sufficiently high amplitudes of alternating magnetic field, respectively. The SAR of iron nanoparticle assembly is shown to exceed significantly that of iron oxide nanoparticles with uniaxial anisotropy at the same amplitudes and frequencies of applied magnetic field. The linear response theory is shown to be valid only at small magnetic field amplitudes, $H_0 \leq 50 - 70$ Oe.

## 1. Introduction

Magnetic nanoparticles are very promising for various applications in biomedicine, such as magnetic resonance imaging, targeted drug delivery, magnetic hyperthermia, etc. [1-6]. However, an assembly of magnetic nanoparticles is a complex physical system. Its properties are determined by various factors, such as the particle magnetic parameters, their average size, external shape, rheological properties of the medium in which the particles are distributed, as well as the concentration of particles in the medium. In addition, the behavior of an assembly significantly depends on the frequency and amplitude of applied alternating (ac) magnetic field.

For application in magnetic hyperthermia, magnetic nanoparticles with high values of specific absorption rate (SAR) in ac magnetic field of moderate amplitude and frequency are required. Experimental measurement of SAR is usually carried out [7-12] for assemblies of nanoparticles distributed in a viscous liquid. It is usually assumed that the magnetic particles possess a uniaxial type of magnetic anisotropy. Meanwhile, nanoparticles with a cubic type of magnetic anisotropy may be very promising for use in magnetic hyperthermia. Indeed, spherical iron nanoparticles have a small single-domain diameter, $D_c = 24$ nm [13], a very high saturation magnetization, $M_s = 1700$ emu/cm$^3$, and a moderate value of the cubic magnetic anisotropy constant, $K_c = 4.6 \times 10^5$ erg/cm$^3$ [14].

In the papers [8,15,16] very high SAR values, of the order of 1 kW/g, were obtained for assemblies of iron nanoparticles distributed in a liquid. High SAR values show also assemblies of biogenic magnetite nanoparticles [17-19], the so called magnetosomes, which are synthesized in nature by magnetotactic bacteria. Magnetosomes have a perfect crystal structure and a narrow particle size distribution, the shape of the nanoparticles being close to spherical. Consequently, such nanoparticles are characterized by a cubic type of magnetic anisotropy.

The SAR calculations [20] for dilute and dense assemblies of magnetosomes distributed in a solid matrix show that nanoparticles with cubic anisotropy are capable of providing sufficiently large SAR values, of the order of 300 - 400 W/g, in an alternating magnetic field of small and moderate amplitude, $H_0 = 50 – 100$ Oe. This fact is extremely important for the practical application of such particles in magnetic hyperthermia. As for the assemblies distributed in a viscous liquid, until now, SAR calculations were carried out [21-23] only for particles with a uniaxial type of magnetic anisotropy.

In this work a theoretical background is developed for the numerical simulation of the properties of dilute assemblies of nanoparticles with cubic magnetic anisotropy distributed in a viscous liquid. Detailed SAR calculations are performed for dilute assemblies of spherical iron nanoparticles with a positive constant of cubic magnetic anisotropy distributed in liquids of different viscosities. The existence of viscous and magnetic modes [21-23] of energy absorption in ac magnetic field is confirmed for nanoparticles with cubic anisotropy, depending on the $H_0$ value. It has been shown that the linear regime of the energy absorption [24] is realized for assemblies of iron nanoparticles only at sufficiently low field amplitudes, $H_0 \leq 50 - 70$ Oe.

The numerical results demonstrate that at the same frequency and amplitude of ac magnetic field iron nanoparticles provide much higher heating efficiency as compared to that of uniaxial nanoparticles [21-23]. The results of the calculations are in qualitative agreement with the experimental data obtained [8, 15] for metallic iron nanoparticles distributed in mesitylene.

## 2. Numerical simulation

The total energy of a spherical single-domain nanoparticle with cubic magnetic anisotropy in ac magnetic field of frequency $f$ and amplitude $H_0$ is

$$W = K_c V \left( (\vec{\alpha}\vec{n}_1)^2 (\vec{\alpha}\vec{n}_2)^2 + (\vec{\alpha}\vec{n}_1)^2 (\vec{\alpha}\vec{n}_3)^2 + (\vec{\alpha}\vec{n}_2)^2 (\vec{\alpha}\vec{n}_3)^2 \right) - M_s V (\vec{\alpha}\vec{H}_0) \cos(\nu t), \qquad (1)$$

where $K_c$ is the cubic magnetic anisotropy constant, $M_s$ is the saturation magnetization, $V$ is the nanoparticle volume and $\nu = 2\pi f$ is the angular frequency. Further, $\boldsymbol{\alpha}$ is the unit magnetization vector and ($\boldsymbol{n}_1$, $\boldsymbol{n}_2$, $\boldsymbol{n}_3$) is the orthogonal set of unit vectors that specifies the spatial orientation of the nanoparticle in the liquid. It is assumed that the set ($\boldsymbol{n}_1$, $\boldsymbol{n}_2$, $\boldsymbol{n}_3$) is rigidly bound, or coincides, with the easy axes of the magnetic anisotropy of a particle that can rotate in a liquid under the influence of thermal fluctuations and the action of ac external magnetic field.

The dynamics of the unit magnetization vector $\boldsymbol{\alpha}$ of a superparamagnetic nanoparticle is described by the stochastic Landau - Lifshitz equation [25-29]

$$\frac{\partial \vec{\alpha}}{\partial t} = -\gamma_1 \left( \vec{\alpha} \times (\vec{H}_{ef} + \vec{H}_{th}) \right) - \kappa \gamma_1 \left( \vec{\alpha} \times (\vec{\alpha} \times (\vec{H}_{ef} + \vec{H}_{th})) \right), \qquad (2)$$

where $\gamma_1 = |\gamma_0|/(1+\kappa^2)$, $\kappa$ is the phenomenological damping constant, $\gamma_0$ being the gyromagnetic ratio. In Eq. (2) $\boldsymbol{H}_{ef}$ is the vector of the effective magnetic field. By definition, it is the variational derivative of the total nanoparticle energy

$$\vec{H}_{ef} = -\frac{\partial W}{V M_s \partial \vec{\alpha}} = \vec{H}_0 \cos(\nu t) - H_k \sum_{i=1}^{3} \left(1 - (\vec{\alpha}\vec{n}_i)^2\right)(\vec{\alpha}\vec{n}_i)\vec{n}_i, \qquad (3)$$

where $H_k = 2K_c/M_s$ is the particle anisotropy field. In Eq. (3) the obvious relation $(\boldsymbol{\alpha n}_1)^2+(\boldsymbol{\alpha n}_2)^2+(\boldsymbol{\alpha n}_3)^2 = 1$ is taken into account. The random thermal magnetic field $\boldsymbol{H}_{th}$ in Eq. (2) is introduced [25] to describe the effect of thermal fluctuations of the unit magnetization vector at a finite temperature $T$. In accordance with the fluctuation-dissipation theorem [25, 28], the components of a random magnetic field have the following statistical properties ($i,j = x,y,z$)

$$\langle H_{th,i}(t) \rangle = 0 ; \qquad \langle H_{th,i}(t) H_{th,j}(t_1) \rangle = \frac{2 k_B T \kappa}{|\gamma_0| M_s V} \delta_{ij} \delta(t-t_1), \qquad (4)$$

where $k_B$ is the Boltzmann constant.

Eqs. (1) - (4) were used earlier [20] to calculate the SAR of nanoparticles with cubic magnetic anisotropy distributed in a solid matrix. However, in the case of a solid matrix, the spatial orientation of each nanoparticle, i. e. the particle frame ($n_1$, $n_2$, $n_3$), is fixed, since in a solid matrix the rotation of nanoparticles is impossible. The behavior of nanoparticles in a liquid is much more complicated, since in this case a nanoparticle can rotate as a whole under the action of mechanical rotational moments in the liquid and the orienting action of an external magnetic field. In this section, a generalization of the procedure [21] is developed to describe the dynamics for nanoparticles with cubic anisotropy capable of rotational motion in a viscous fluid.

The kinematic equations of motion for the particle frame ($n_1$, $n_2$, $n_3$) are given by

$$\frac{d\vec{n}_1}{dt} = \vec{\omega} \times \vec{n}_1; \qquad \frac{d\vec{n}_2}{dt} = \vec{\omega} \times \vec{n}_2; \qquad \frac{d\vec{n}_3}{dt} = \vec{\omega} \times \vec{n}_3, \qquad (5)$$

where the vector $\omega$ is the mechanical angular frequency of rotation of the nanoparticle as a whole. In a viscous liquid the rotational motion of a spherical nanoparticle is described by the corresponding stochastic equation

$$I\frac{d\vec{\omega}}{dt} + \xi\vec{\omega} = \vec{N}_m + \vec{N}_{th}. \qquad (6)$$

Here $I$ is the moment of inertia of a spherical nanoparticle, $\xi = 6\eta V$ is the coefficient of friction of a rotating nanoparticle in a viscous liquid, obtained in the Stokes approximation for the case of small Reynolds numbers [30], $\eta$ is the liquid viscosity. In Eq. (6) $N_m$ is a regular mechanical moment applied to a particle, and $N_{th}$ is a fluctuating rotational moment, which leads to rotational Brownian motion of a nanoparticle in a liquid even in the absence of an external magnetic field and not associated with the magnetic properties of the nanoparticle.

The regular mechanical moment $N_m$ in Eq. (6), related with the presence of magnetic degrees of freedom of the particle, is calculated by the equation

$$\vec{N}_m = \sum_{i=1}^{3}\left(\frac{\partial W}{\partial \vec{n}_i} \times \vec{n}_i\right) = 2K_c V \sum_{i=1}^{3}(\vec{\alpha}\vec{n}_i)\left(1-(\vec{\alpha}\vec{n}_i)^2\right)(\vec{\alpha}\times\vec{n}_i). \qquad (7)$$

This equation shows that the motion of the magnetic vector through the energy of magnetocrystalline anisotropy affects the spatial orientation of the trihedron ($n_1$, $n_2$, $n_3$). In accordance with the fluctuation-dissipation theorem [28], the components of the fluctuating moment $N_{th}$ have the following statistical properties ($i,j = x,y,z$)

$$\langle N_{th,i}(t)\rangle = 0; \qquad \langle N_{th,i}(t)N_{th,j}(t_1)\rangle = 2k_B T\xi\delta_{ij}\delta(t-t_1). \qquad (8)$$

Note that due to the assumed orthonormality of the vectors ($n_1$, $n_2$, $n_3$), their dynamics can be described using two arbitrarily chosen kinematic Eqs. (5), setting, for example,

$$\frac{d\vec{n}_1}{dt} = \vec{\omega} \times \vec{n}_1; \qquad \frac{d\vec{n}_2}{dt} = \vec{\omega} \times \vec{n}_2; \qquad \vec{n}_3 = \vec{n}_1 \times \vec{n}_2, \qquad (9)$$

and assuming that at the initial moment of time the unit vectors $n_1$ and $n_2$ are orthogonal, so that $n_1(0)n_2(0) = 0$. Indeed, in this case, the orthogonality of the vectors $n_1$ and $n_2$ is preserved for all times, since

$$\frac{d}{dt}(\vec{n}_1\vec{n}_2) = \vec{n}_1\frac{d\vec{n}_2}{dt} + \vec{n}_2\frac{d\vec{n}_1}{dt} = \vec{n}_1(\vec{\omega}\times\vec{n}_2) + \vec{n}_2(\vec{\omega}\times\vec{n}_1) = 0.$$

Further, due to the small size of the magnetic nanoparticle, it is possible to neglect its moment of inertia in Eq. (6), assuming $I \approx 0$. Then, taking into account Eqs. (6), (7), one obtains a closed system of equations for determining the dynamics of the basis vectors in the form

$$\frac{d\vec{n}_1}{dt} = -G(\vec{\alpha}\vec{n}_1)\left(\vec{n}_1(\vec{\alpha}\vec{n}_1)^3 + \vec{n}_2(\vec{\alpha}\vec{n}_2)^3 + \vec{n}_3(\vec{\alpha}\vec{n}_3)^3 - \vec{\alpha}(\vec{\alpha}\vec{n}_1)^2\right) - \frac{1}{\xi}(\vec{n}_1 \times \vec{N}_{th});$$

$$\frac{d\vec{n}_2}{dt} = -G(\vec{\alpha}\vec{n}_2)\left(\vec{n}_1(\vec{\alpha}\vec{n}_1)^3 + \vec{n}_2(\vec{\alpha}\vec{n}_2)^3 + \vec{n}_3(\vec{\alpha}\vec{n}_3)^3 - \vec{\alpha}(\vec{\alpha}\vec{n}_2)^2\right) - \frac{1}{\xi}\left(\vec{n}_2 \times \vec{N}_{th}\right); \quad (10)$$
$$\vec{n}_3 = \vec{n}_1 \times \vec{n}_2,$$

where the coefficient $G = 2K_cV/\xi = K_c/3\eta$ does not depend on the particle radius. It is assumed that the dynamics of the unit magnetization vector $\vec{\alpha}$ obeys Eq. (2).

Eqs. (2) and (10), together with Eqs. (4), (8), form a closed system of equations for describing the behavior of a magnetic nanoparticle with cubic anisotropy in a viscous liquid in applied magnetic field. These equations are solved in this work for the case of a dilute assembly of iron nanoparticles using the well-known Heun algorithm [26–27]. The procedure for calculating the SAR of an assembly in ac magnetic field is described in detail in [21].

## 3. Results and discussion

### 3.1. *SAR of dilute assembly of iron nanoparticles in a viscous liquid*

The SAR calculations for a dilute assembly of spherical iron nanoparticles were carried out for liquids of different viscosities, $\eta$ = 0.01 - 0.1 g/cm/s, in the range of particle diameters $D$ = 6 - 24 nm, not exceeding the single-domain diameter of iron nanoparticles, $D_c$ = 24 nm. The ac magnetic field amplitude varied in the range $H_0$ = 30 – 300 Oe, the field frequency $f$ = 200 - 300 kHz, the temperature of the system $T$ = 300 K, and the magnetic damping constant $\kappa$ = 0.5.

As Fig. 1a shows, the dependence of the SAR on the particle diameter at low field amplitudes, $H_0$ = 50 Oe, has a characteristic bell-shaped form. An increase in the liquid viscosity has practically no effect on the SAR of particles with diameters $D \leq 19$ nm, however, it leads to a sharp drop in the SAR at $D \geq 20$ nm. Nevertheless, for particles of optimal diameters, $D$ = 18 - 20 nm, the SAR of the assembly exceeds the value of 300 W/g at $H_0$ = 50 Oe and frequency $f$ = 200 kHz, regardless of the liquid viscosity. At a viscosity $\eta$ > 0.1 g/cm/s, the dependence of the SAR on the particle diameter practically coincides with the SAR of an assembly of nanoparticles hindered in a nonmagnetic matrix [20].

According to Fig.1b, with an increase in the field amplitude, $H_0 \geq 100$ Oe, the SAR of the assembly increases monotonically with an increase in the particle diameter and exceeds 1 kW/g for nanoparticles with diameters $D \geq 20$ nm. Such high values of the SAR of iron nanoparticles at rather moderate field amplitudes are explained, most probably, by the large value of the saturation magnetization of iron nanoparticles.

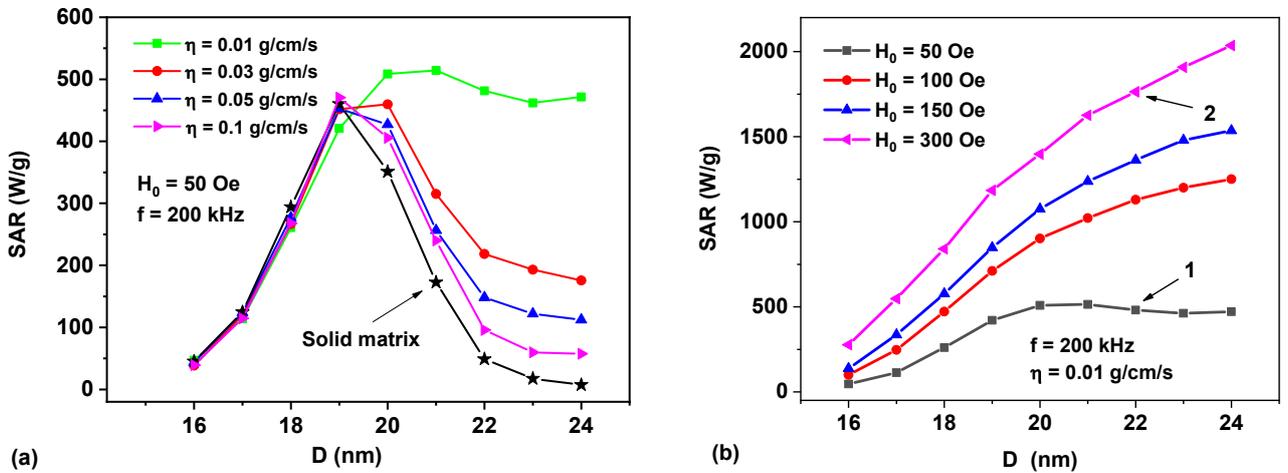

Fig. 1. a) Comparison of SAR of dilute assemblies of iron nanoparticles distributed in a solid matrix and in liquids with a viscosity $\eta$ = 0.01; 0.03; 0.05 and 0.1 g/cm/s, respectively. b) the dependence of the SAR on the ac magnetic field amplitude at a fixed frequency and liquid viscosity.

## 3.2. Magnetization reversal modes

It has been theoretically shown [21-23] that for an assembly of uniaxial nanoparticles in a viscous liquid, the difference in the behavior of the SAR assembly at small and relatively large amplitudes of ac magnetic field is associated with the implementation of viscous and magnetic modes of particle magnetization reversal. At low field amplitudes, in the steady-state viscous mode, the unit magnetization vector $\alpha$ and the particle director $n$ move approximately in unison. On the contrary, in the developed magnetic mode, the unit magnetization vector of the particle jumps between the equivalent magnetic potential wells with the field frequency, while the director of the particle experiences only small oscillations about the magnetic field direction during the jump of the vector $\alpha$.

For iron nanoparticles with cubic anisotropy, similar modes of motion of the vectors $\alpha$ and the frame ($n_1$, $n_2$, $n_3$) were found in this work. The main difference between these dynamic processes is that for a spherical iron nanoparticle there are three equivalent directors and, accordingly, in a relatively weak ac magnetic field, there are six equivalent potential wells, between which the magnetic vector can jump during its motion. The black solid curve in Fig. 2a shows the dynamics of the projection of the unit magnetization vector $\alpha_x$, parallel to the magnetic field direction, in the viscous mode over several periods of field variation for randomly selected particle of the assembly. The red dots in this figure show the dynamics of the $x$-component of one of the particle directors, $n_{2x}$. As Fig. 2a shows, in the steady-state viscous mode, the $x$-projections of the vectors $\alpha$ and $n_2$ can move in unison for many periods of the field, although over time the vector $n_2$ can change to equivalent vectors, $n_1$ or $n_3$, due to random sharp rotations of the particle.

However, as Fig. 2b shows, the dynamics of the vectors changes significantly with an increase in the amplitude of the alternating magnetic field. So, in the developed magnetic mode, at $H_0 = 300$ Oe, one of the equivalent directors, in the given case the vector $n_1$, remains approximately parallel to the field direction, while the magnetic vector jumps between the values $\alpha_x = \pm 1$ with the frequency of the alternating field. Calculations show that in a developed magnetic mode, over time, the vectors $n_1$, $n_2$ and $n_3$ can also change places. But at least one of these vectors remains parallel to the ac magnetic field direction, while the movement of other vectors of the basis remains chaotic.

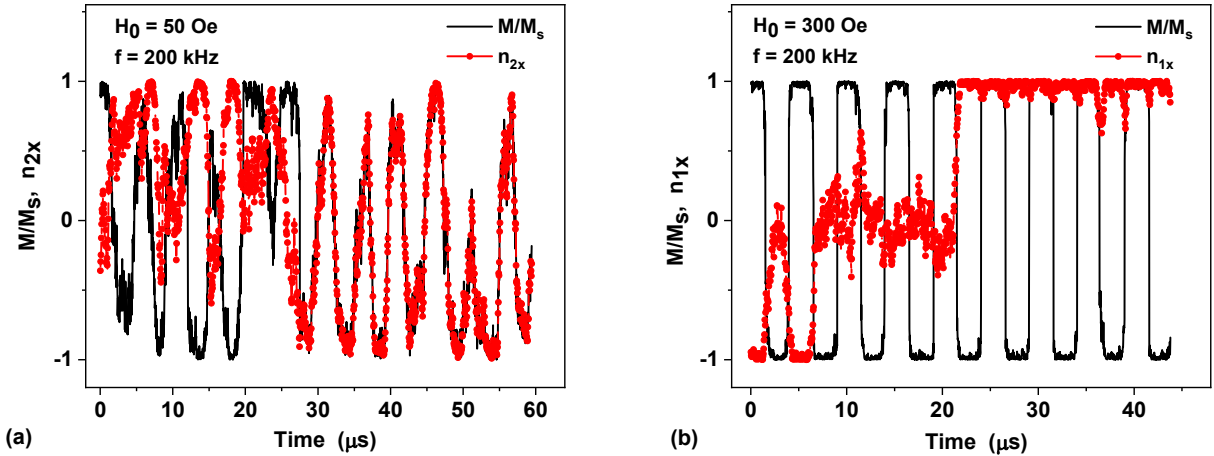

Fig. 2. The magnetization reversal modes of iron nanoparticles in a viscous liquid: a) viscous mode, $H_0 = 50$ Oe, $f = 200$ kHz; b) magnetic mode, $H_0 = 300$ Oe, $f = 200$ kHz. The viscosity of the liquid is $\eta = 0.01$ g/cm/s, the particle diameter is $D = 22$ nm. The dynamics of vectors shown in Fig. 2a and 2b corresponds to points 1 and 2 marked in Fig. 1b.

## 3.3. Linear response theory

In the work of Rosensweig [24] the SAR of a dilute assembly of nanoparticles in a liquid was calculated in a linear approximation

$$SAR = \pi \chi_0 \frac{\nu \tau_{ef}}{1+(\nu \tau_{ef})^2} \frac{f}{\rho} H_0^2, \qquad (11)$$

where $\chi_0 = M_s^2 V/3k_B T$ is the initial magnetic susceptibility of the assembly that does not depend on the type of particle magnetic anisotropy, and $\tau_{ef}$ is effective Schliomis relaxation time [31]

$$\tau_{ef} = \frac{\tau_B \tau_N}{\tau_B + \tau_N}. \qquad (12)$$

The latter was postulated under the assumption that the orientational and magnetic relaxation processes in a dilute assembly of nanoparticles in a liquid are independent.

In Eq. (12) $\tau_B = 3\eta V_\eta /k_B T$ is the characteristic time of Brownian orientational relaxation of nanoparticles in a liquid [32], $V_\eta$ is the effective volume of a nanoparticle. This value takes into account the possible presence of a nonmagnetic layer on the particle surface. In this work we assume for simplicity $V_\eta = V$. Moreover, $\tau_N$ is the Neél relaxation time of the average magnetic moment for assembly of immobile superparamagnetic particles. For the case of spherical iron nanoparticles with a positive cubic anisotropy constant an analytical estimate of $\tau_N$ was obtained [33] in the limit of a sufficiently large reduced energy barrier, $K_c V/4k_B T >> 1$

$$\tau_N = \frac{1}{f_{0c}} \exp\left(\frac{K_c V}{4k_B T}\right). \qquad (13a)$$

In our case of intermediate to high damping limit, $\kappa = 0.5$, for the attempt frequency $f_{0c}$ in Eq. (13a) one should use the expression

$$f_{0c} = \frac{4\gamma_1 K_c (\kappa+2q)}{\pi M_s} \sqrt{\frac{2(\kappa+2q)}{\kappa(1+12q^2)+16q^3}}; \qquad q = \frac{\sqrt{9\kappa^2+8}-3\kappa}{8}. \qquad (13b)$$

It has recently been shown [23] that Shliomis's hypothesis, Eq. (12), adequately describes the process of magnetic relaxation of an assembly of uniaxial nanoparticles in liquids of different viscosities. It is of interest to check the validity of equations (12), (13) for the case of nanoparticles with cubic anisotropy. For this purpose, let us prepare the oriented initial state of the assembly at the time $t = 0$, so that the unit magnetization vectors of each particle and one of its directors are parallel to the $z$-axis of the Cartesian coordinates, $\mathbf{n}_{1i} = \boldsymbol{\alpha}_i = (0,0,1)$, $i = 1, 2, .., N_p$. Without loss of generality one can assume that such a director is the vector $\mathbf{n}_1$. The time evolution of the assembly for $t > 0$ for a given viscosity $\eta$ and temperature $T$ can be studied by solving the system of stochastic equations of motion for particle directors and unit magnetization vectors obtained in Section 2. Average reduced magnetic moment of the assembly and average moment of the set of vectors $\mathbf{n}_{1i}$ are calculated by the formulas

$$\langle \vec{m}(t) \rangle = \frac{1}{N_p} \sum_{i=1}^{N} \vec{\alpha}_i; \qquad \langle \vec{n}(t) \rangle = \frac{1}{N_p} \sum_{i=1}^{N} \vec{n}_{1i}. \qquad (14)$$

The latter vector characterizes the change in the total spatial orientation of the particles of the assembly over time. Obviously, as a result of relaxation of the dilute assembly of superparamagnetic nanoparticles to the equilibrium state, the average value of both vectors should be equal to zero, $<\mathbf{m}> = <\mathbf{n}> = 0$.

The relaxation process of a dilute assembly of iron nanoparticles in a liquid was calculated for particles of different diameters for assemblies with a sufficiently large number of particles, $N_p = 1000$. The numerical time step was chosen equal to 1/30 of the characteristic precession time $T_p$ of the particle unit magnetization vectors. This guarantees [29] sufficient accuracy of the numerical calculations performed.

As Figs. 3a, 3b show, for spherical iron nanoparticles with diameters $D \leq 18$ nm, the relaxation of the average magnetization of the assembly (filled circles) occurs mainly due to the overcoming of energy barriers between the potential wells of particles by unit magnetization vectors, since for particles of relatively small diameters the relaxation time $\tau_N \ll \tau_B$. At the same time, according to Fig. 3c for particles with diameters $D \geq 20$ nm the magnetic relaxation of the assembly occurs mostly due to the rotation of nanoparticles in the liquid, since in this case $\tau_N \gg \tau_B$. Note that in all the cases considered, the orientational relaxation of the assembly (squares) is well described by the Debye relaxation curve (triangles), which represents the orientational relaxation of nonmagnetic nanoparticles in a liquid. Moreover, the process of magnetic relaxation of the iron nanoparticle assembly is quite accurately described by exponential curves with the relaxation time $\tau_{ef}$ (solid curves in Fig. 3), calculated according to Eqs. (12), (13). Similar results were obtained for iron nanoparticles in liquids of various viscosities. Consequently, Shliomis's hypothesis (12) is fully justified for the case of relaxation of the magnetic moment of a dilute assembly of iron nanoparticles in a viscous liquid in the absence of external magnetic field. It can be assumed that it remains valid also in a weak magnetic field, which justifies the use of the value $\tau_{ef}$ in the Eq. (11). On the other hand, it is important to clarify the range of the ac magnetic field amplitudes, in which the Eq. (11) itself is valid.

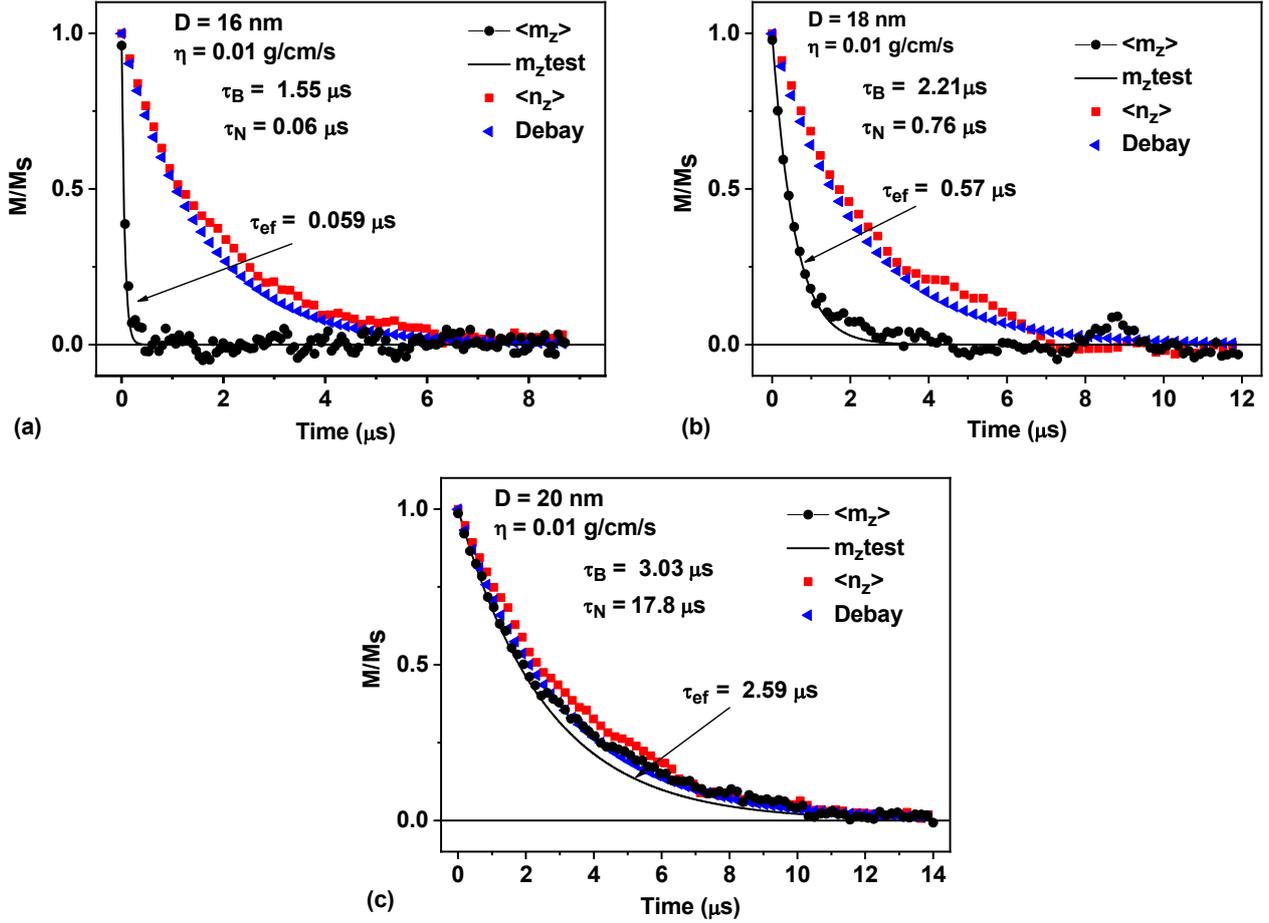

Fig. 3. Relaxation of an assembly of iron nanoparticles of various diameters in a liquid with viscosity $\eta$ = 0.01 g/cm/s: a) $D$ = 16 nm, b) $D$ = 18 nm, c) $D$ = 20 nm. Filled circles show the relaxation process of the assembly magnetization, squares show the process of orientational relaxation, Debye relaxation is shown by triangles, the solid curves show exponential decay of magnetization with the corresponding relaxation time $\tau_{ef}$.

For this purpose, the results of numerical simulation of the SAR of a dilute assembly of iron nanoparticles are compared in Fig. 4 with the predictions of the linear response theory (LRT), Eq. (11). As Fig. 4 shows, for iron nanoparticles in all the cases considered the LRT is fulfilled only in the range of small amplitudes of the ac magnetic field, $H_0 \leq 50 - 70$ Oe. With an increase in the field amplitude, the numerical results for the SAR turn out to be significantly less than the LRT predictions, Eq. (11).

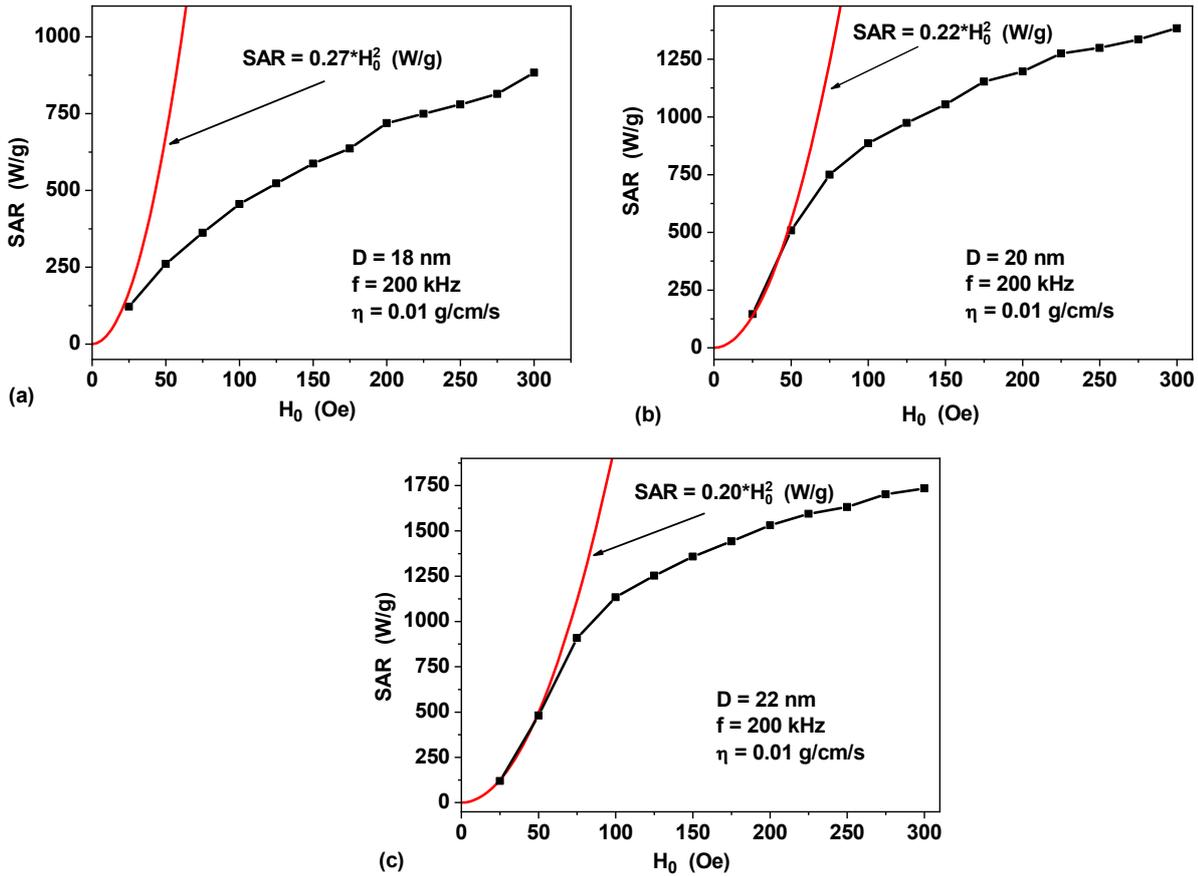

Fig. 4. Comparison of the LRT (solid curves) with the results of the numerical calculation of the SAR (filled squares) for spherical iron nanoparticles of various diameters: a) $D$ = 18 nm, b) $D$ = 20 nm, c) $D$ = 22 nm.

### 3.4. *SAR comparison for iron and iron oxide nanoparticles*

It is interesting to compare the numerical data for the SAR of dilute assembly of iron nanoparticles with that of an assembly of uniaxial iron oxide nanoparticles [23] with typical saturation magnetization $M_s$ = 350 emu/cm$^3$ and uniaxial anisotropy constant $K_1$ = 10$^5$ erg/cm$^3$. As Fig. 5a shows at the same frequency and amplitude of ac magnetic field in the range of diameters $D$ = 16 - 24 nm the SAR of assembly of iron nanoparticles significantly exceeds that of iron oxide nanoparticles. This demonstrates one again the very high heating efficiency of assembly of spherical iron nanoparticles.

For comparison with the experimental results [8,15], we also calculated SAR for dilute assemblies of iron nanoparticles of various average diameters at ac magnetic field frequency $f$ = 300 kHz, and a liquid viscosity $\eta$ = 0.01 g/cm/s, close to the mesitylene viscosity. As Fig. 5b shows, the SAR values of the assembly grow rather rapidly with an increase in the amplitude of the ac magnetic field and an increase in the particle diameter. Similar tendencies were also observed in experiments [8, 15], although for a number of samples in [15] a more complex behavior of SAR was found, which may be associated with the formation of dipole-coupled one-dimensional chains of nanoparticles in a liquid. This issue requires special consideration. On the whole, the numerical calculations of SAR are in qualitative agreement with the experimental results. For example, for particles with diameter $D$ = 16 nm at $f$ = 300 kHz and $H_0$ = 400 - 500 Oe the SAR values of the order of 1000 - 1100 W/g were obtained [8], which is in agreement with the data presented in Fig. 5b.

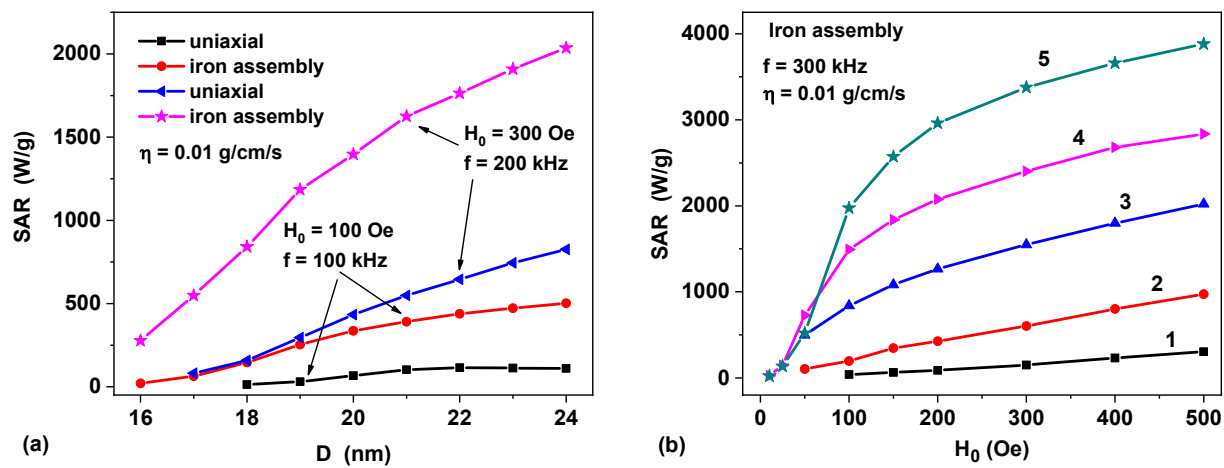

Fig. 5. a) Comparison of the heating efficiency of dilute assemblies of iron nanoparticles with cubic anisotropy and uniaxial iron oxide nanoparticles in the range of diameters $D$ = 16 - 24 nm. b) SAR of diluted assembly of iron nanoparticles of various diameters: 1) $D$ = 14 nm, 2) $D$ = 16 nm, 3) $D$ = 18 nm, 4) $D$ = 20 nm, 5) $D$ = 24 nm.

## 4. Conclusions

It was shown experimentally [15 - 16] that metallic iron nanoparticles of cubic and quasi-spherical shape with diameters $D$ = 6 - 27 nm exhibit very high SAR values being distributed in a viscous liquid. The use of such nanoparticles would be very promising in magnetic hyperthermia. However, it is necessary to protect metallic iron nanoparticles from the aggressive action of the biological environment, at least for the duration of treatment in magnetic hyperthermia. It can be made by creating silica coated iron nanoparticles [16], or using other, more reliable non magnetic particle shells of sufficient thickness.

      In this work, we performed a numerical calculation of SAR for dilute assemblies of spherical iron nanoparticles with a cubic type of magnetic anisotropy, and confirmed the high efficiency of heat production by iron nanoparticles in a viscous liquid even at moderate frequencies and amplitudes of ac magnetic field. In calculations it was taken into account that, in contrast to magnetic nanoparticles with uniaxial anisotropy [21-23], the spatial orientation of nanoparticles with cubic anisotropy is described by three orthonormal directors. This requires a more general theoretical description. It has been shown that depending on the amplitude of the ac magnetic field, for nanoparticles with cubic anisotropy viscous and magnetic modes of particle magnetization reversal are realized similar to the case uniaxial nanoparticles [21,34,35]. The range of applicability of the linear response theory [24] was investigated by means of comparison with numerical simulation data. The latter seem to be in qualitative agreement with the experimental data [8,15].

## Acknowledgements

E.G. gratefully acknowledges the financial support of the RFBR grant # 20-32-90085. R.R gratefully acknowledges the financial support of the Ministry of Science and Higher Education of the Russian Federation in the framework of Increase Competitiveness Program of NUST «MISIS», contract № K2A-2019-034.